\documentclass[12pt]{article}
\usepackage{epsfig}

\input psfig

\def\xslide#1#2#3#4#5#6#7{\centerline{\psfig
{figure=#1,height=#2,bbllx=#3bp,bblly=#4bp,bburx=#5bp,bbury=#6bp,width=#7,clip=}}}

\newcommand{\be }{\begin{equation}}
\newcommand{\ee}{\end{equation}} 
\newcommand{\ba}{\begin{eqnarray}}
\newcommand{\ea}{\end{eqnarray}} 
\newcommand{\pp}{$\pi\pi$ }
\newcommand{\pipi}{$\pi^+\pi^-$ }
 
\newcommand{\fo}{$f_0(980)$ }

\newcommand{\epw}{$f_0(1400)$ }

\newcommand{\kk}{$K\overline{K}$ }
\newcommand{\epsig}{$f_0(500)$ }
\newcommand{\eq}{\begin{equation}}
\newcommand{\qe}{\end{equation}}

\newcommand{\roro}{$\sigma\sigma$ }

\begin{document}

\title{Peculiarities in multichannel interaction amplitudes for meson-meson
scattering and scalar meson spectroscopy}
\author{R.  Kami\'nski$^a$, L. Le\'sniak$^a$ and B. Loiseau$^b$ \\
\small{
$^a$Henryk Niewodnicza\'nski Institute of Nuclear Physics,}\\ 
\small{PL 31-342 Krak\'ow, Poland}\\
\small{$^b$LPNHE/LPTPE Universit\'e P. et M. Curie, 4, Place Jussieu,}\\
\small{75252 Paris CEDEX 05, France}}

\maketitle

\begin{abstract}
Interactions in coupled channels $\pi \pi$, $K \bar K$ and 
an effective $2\pi 2\pi$ in scalar-isoscalar wave have been analysed. 
Influence of interchannel couplings on analytical structure of
multichannel interaction amplitudes has been studied.  
Interplay of $S$-matrix zeroes and poles and their relation with parameters
of scalar resonances has been investigated. 
\end{abstract} 
 
 
\vspace{0.2cm} 
 
Structure and dynamics of scalar mesons is still not clear and is a 
subject of many theoretical and experimental efforts
\cite{pdg98}.  
Their multichannel decays and possible interference with glueball states 
below 2 GeV make results of analyses very model-dependent.
In our model we use separable potentials 
\cite{kll2, ll96}
for three coupled channels:
\pp, \kk and \roro (an effective $2\pi 2\pi$) in $S$-wave with isospin 0.
We also do not assume the existence of any scalar mesons before 
fitting experimental data.
%and do not make any assumptions about analytical structure of the
%scattering amplitudes.
As experimental data for the \pipi interactions we use "down-flat" and "up-flat" 
solutions of
\cite{klr}
in the effective two-pion mass
$m_{\pi\pi}$ range from 600 MeV to 1600 MeV.
Below 600 MeV we use the \pp data described in 
\cite{kll2} and 
 the \kk channel data of \cite{cohen}.

\begin{figure}[ht!] 
\xslide{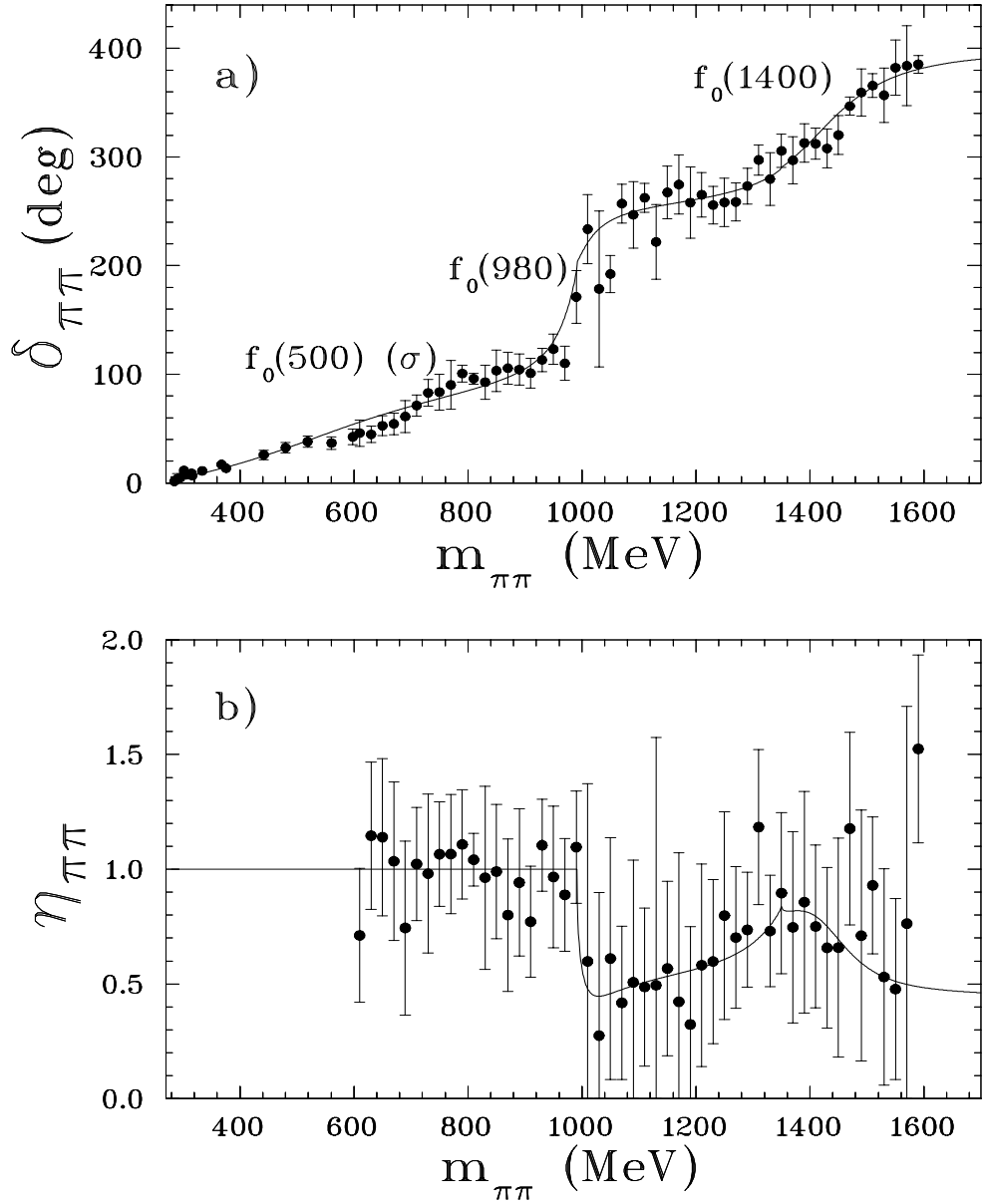}{16cm}{45}{50}{550}{716}{13cm}

\vspace{0.4cm}

\caption{Energy dependence of {\bf a)} \pp phase shifts and
 {\bf b)} \pp inelasticity for the solution B of [2].
 Experimental data above $m_{\pi\pi}=600$ MeV correspond to the "down-flat"
 solution from [4].}
\label{phase_eta}
\end{figure}

We solve a system of three Lipp\-mann-Schwinger coupled equations and find the 
Jost functions $D(k_1,k_2,k_3)$. 
Then we construct $S$-matrix elements 
and express them in terms of phase shifts and inelasticities which we
fit to available experimental data. 
In this way we determine 14 free parameters 
 of our model, finding two solutions to the "up-flat" data and four solutions 
 to the "down-flat" data. We obtain $\chi^2$ values smaller than 116 for 102 
 points.
An example of the energy dependence of the \pp phase shifts and inelasticities 
for one of our solutions is shown in Fig. \ref{phase_eta}.

In order to investigate a role played by the interchannel interactions in the
scalar meson dynamics we first study the analytical structure of the
$S$-matrix elements in the fully decoupled case where all the interchannel
couplings are equal to zero.
In such a case one can recognize in which channel particular poles are created.
When the interchannel couplings are switched on, all the poles change positions
and split into $2^{n-1}$ ($n$ is the number of channels) poles lying on 
different sheets labeled by the signs
of the imaginary parts of the complex momenta in various channels.
Both in the fully decoupled and in the coupled cases all the diagonal $S$-matrix 
elements can be expressed as ratios of two
Jost functions. The denominators of these matrix elements are the same but their
numerators are different.
Therefore the positions of poles are common for all 
the $S$-matrix elements but the positions of zeroes depend on the chosen
channel. Only poles and zeroes which lie close enough to 
the physical region have a significant influence on the experimental 
phase shifts and  inelasticities.
We then study poles and zeroes nearby the physical regions and relate scalar 
resonances to them.

\begin{figure}[ht!] 
\xslide{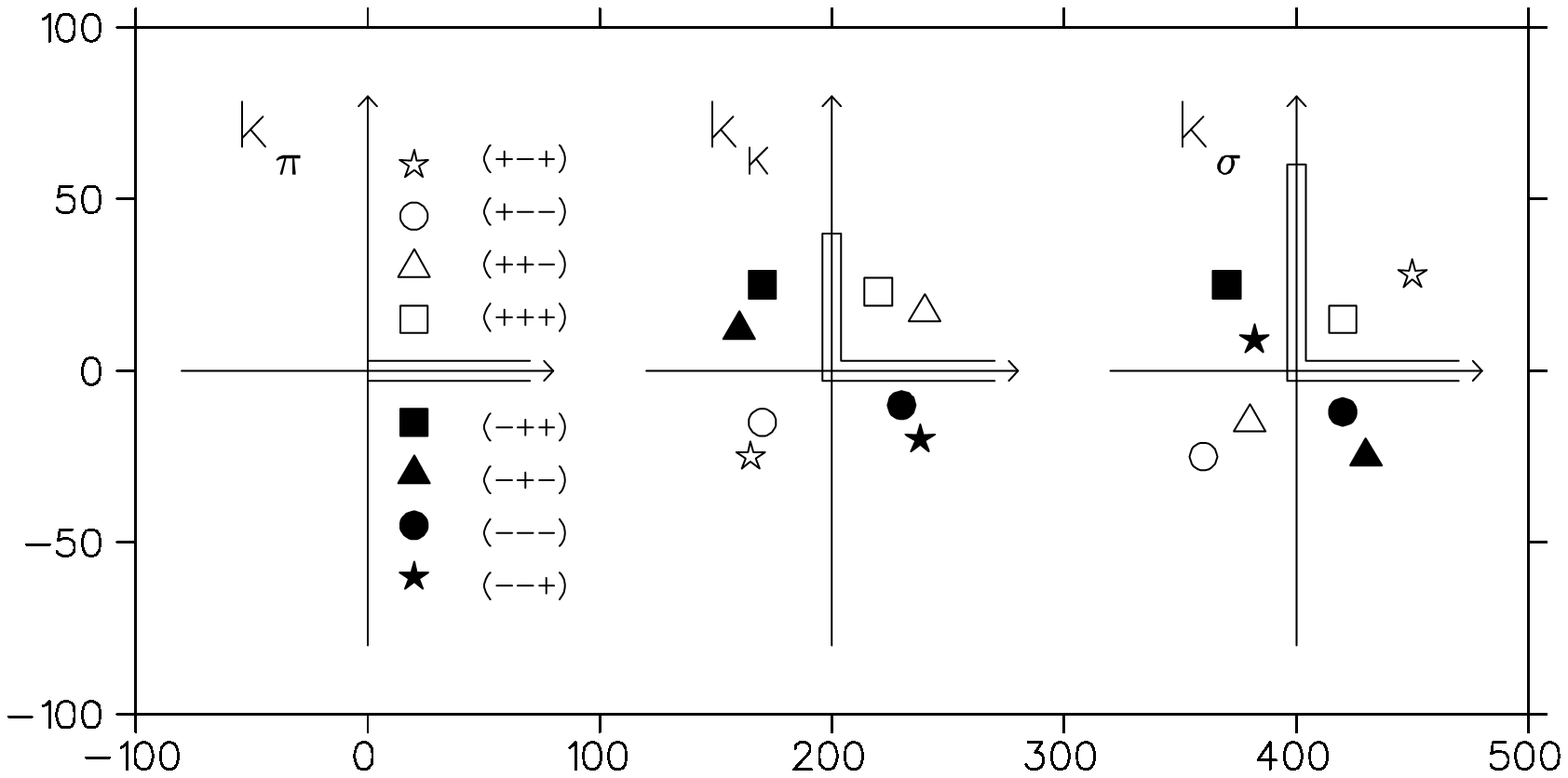}{7cm}{128}{512}{536}{720}{12.5cm}

%\vspace{0.4cm}

\caption{Complex momentum planes with schematic positions of the $S_{\pi\pi}$
zeroes (empty symbols) and poles (full symbols). 
Sheet positions are given on the $k_{\pi}$ plane. 
For clarity, positions of the symmetric poles and zeroes with 
respect to the imaginary axes are not plotted. 
Double lines denote physical regions in the particular channels.}
\label{3complex}
\end{figure}

In Fig. \ref{3complex} we show schematic positions of some poles and zeroes of
the $S_{\pi\pi}$-matrix element on various planes of complex momenta. However,
not all the poles and zeroes are equally important.
One can see that below the \kk threshold only a pole on sheet $(-++)$ and a zero 
on sheet $(+++)$ lie near physical region in all the complex momenta planes.  
Therefore, they play the most important role in that region 
and can be related to resonances $f_0(500)$ (or $\sigma$) and $f_0(980)$.
The second resonance is still interpreted either as an ordinary
$q\bar q$ state or as a \kk quasi-bound state.  
It was already pointed out in \cite{kll2} that
the nature of $f_0(980)$ can be investigated by watching positions of the poles 
while the interchannel couplings go to zero.
In two of our solutions we find that in the fully decoupled case the $f_0(980)$
poles are located
on the positive part of imaginary $k_K$ axis forming the \kk bound state.
In two other solutions the corresponding poles lie below the real $k_K$ axis
thus exhibiting the ordinary resonance nature of $f_0(980)$. In order to resolve
this ambiguity one needs, however, new and very precise experimental data on the 
\kk interaction amplitudes near threshold. 

Near the \roro threshold we find four poles but in Table \ref{resonances} only 
two of them are listed.
The two poles on sheets $(---)$ and $(--+)$ play a dominant
role in the \roro channel. 
Different positions of those poles may lead to their interpretation 
as two distinct resonances (with different masses, widths, 
branching ratios etc.). 
We know, however, that in three of our solutions they come from the "bare" \pp 
resonance which splits into four states due to the interchannel couplings.
In one case, namely in the solution B, two poles near 1400 MeV have their origin
in the \pp channel and the other two in the \roro channel. Since the masses of
resonances determined by positions of the poles
are similar, in Table \ref{resonances} we use one common name for
them: $f_0(1400)$.
Note that our two $f_0(1400)$ poles can be compared to the wide $f_0(1370)$
and the narrow $f_0(1500)$ listed in
\cite{pdg98}.
\begin{table}[h!]
\centering
\caption{Average masses and widths of resonances \epsig, \fo and 
\epw found in our solutions A, B, E and F (fits to "down-flat" data sets). 
Errors represent the maximum departure from the average.} 

\vspace{0.3cm}

\begin{tabular}{|c|c|c|c|}
\hline
resonance & mass (MeV) & width (MeV) & sheet \\
\hline 
\epsig or $\sigma$ & $523 \pm 12$ & $518 \pm 14$ & $-++$ \\
\hline
\fo & $991 \pm 3$ & $71 \pm 14$ &  $-++$ \\
\hline
& $1406 \pm 19$ & $160 \pm 12$ & $---$ \\
\epw & $1447  \pm 27$ & $108 \pm 46$ & $--+$ \\
\hline 
\end{tabular}
\label{resonances}
\end{table}
%%%%%%%%%%%%%%%%%%%%%%%%%%%%%%%%%%%%%%%%%%%%%%%%%%%%%%%%%%%%%%%%%

%As can be seen parameters of both the $f_0(500)$ and $f_0(980)$  agree well
%with those of Particle Data Tables 
%\cite{pdg98}, 
%however,   
%in the energy range 1300-1500 MeV one can find there 
%two near lying states: a wider $f_0(1370)$ and a narrower $f_0(1500)$.

Other properties of resonances listed in Table \ref{resonances} 
(e.g. branching ratios) together with a
discussion of the limited applicability of the Breit-Wigner approach can be found in
our paper 
\cite{kll2}.
%We also present there parametrisation of amplitudes with the use of known 
%positions of only single pole and zero related to a particular resonance.
%This parametrisation applied for example in vicinity of $f_0(1400)$ allows to 
%calculate for example phase shifts with good accuracy with those of the full our model. 

%%%%%%%%%%%%%%%%%%%%%%%%%%%%%%%%%%%%%%%%%%%%%%%%%%%%%%%%%%%%%%%%%%%%%%%%%%%%%%%%%%%%%%

%%%%%%%%%%%%%%%%%%%%%%%%%%%%%%%%%%%%%%%%%%%%%%%%%%%%%%%%%%%%%%%%%%%%%%
\end{document}